\documentclass[conference]{IEEEtran}
\usepackage{cite}
\usepackage{graphicx}
\usepackage{psfrag}
\usepackage{subfigure}
\usepackage{url}
\usepackage{stfloats}
\usepackage{amsmath, amssymb, bm}
\usepackage[pointedenum]{paralist}
\usepackage{array}
\usepackage[margin=0.64in]{geometry}
\DeclareGraphicsExtensions{.eps,.ps}
\graphicspath{{./eps/}}

\begin{document}

\title{On Optimum End-to-End Distortion in \\ Wideband MIMO Systems}

\author{\IEEEauthorblockN{Jinhui Chen\IEEEauthorrefmark{1}, Dirk T. M. Slock\IEEEauthorrefmark{2}}
\IEEEauthorblockA{\IEEEauthorrefmark{1}Research \& Innovation Center, Alcatel-Lucent Shanghai Bell \\ Email: Jinhui.Chen@alcatel-sbell.com.cn\\}
\IEEEauthorblockA{\IEEEauthorrefmark{2}Department of Mobile Communications, EURECOM \\Email: Dirk.Slock@eurecom.fr}
}
\maketitle

\begin{abstract}
This paper presents the impact of frequency diversity on the optimum expected end-to-end distortion (EED) in an outage-free wideband multiple-input multiple-output (MIMO) system. We provide the closed-form expression of optimum asymptotic expected EED comprised of the optimum distortion exponent and the multiplicative optimum distortion factor for high signal-to-noise ratio (SNR). It is shown that frequency diversity can improve EED though it has no effect on ergodic capacity. The improvement becomes slight when the frequency diversity order is greater than a certain number. The lower bounds related to infinite frequency diversity are derived. The results for outage-free systems are the bounds for outage-suffering systems and they are instructive for system design.
\end{abstract}

\IEEEpeerreviewmaketitle

\section{Introduction}

Generally, in analog source transmission, the end-to-end distortion (EED), \emph{i.e.}, the distortion in the recovered analog source at the receiver, is the primary metric for measuring the performance of an entire transmission system including source and channel coding. There are various scenarios with different source-to-channel bandwidth ratios (SCBR), $W_s/W_c$. For instance, a video transmission system would be at a high SCBR  whereas the channel-parameter feedback procedure would be at a low SCBR.

Caire-Narayanan \cite{caire_allerton05, caire_it07} and Gunduz-Erkip \cite{gunduz_itw06} derived the optimum distortion exponent in the optimum expected EED for outage-free  systems over spatially uncorrelated MIMO block-fading flat channels. We derived the optimum asymptotic expected EED for high SNR, comprised of the optimum distortion exponent and the multiplicative optimum distortion factor for both cases of spatially uncorrelated and correlated block-fading flat channels \cite{chen_icc08, chen_globecom08, chen_eurasip09}. Concurrently, Tuninetti \emph{et al.} also showed that the spatial correlation degrades the achievable expected EED in power-offset, \emph{i.e.}, distortion factor, but not affects the distortion exponent \cite{tuninetti_icc09}.

A wideband channel can be considered as a set of parallel narrow-band flat subchannels \cite{tse, goldsmith}. The approaches of multi-carrier aggregation, OFDM modulation with subcarrier interleaving and spread-spectrum have been employed to harness the frequency diversity. For an outage-suffering system, it has been demonstrated that the frequency diversity can mitigate the outage probability.

In this paper, based on our preceding results on flat channels \cite{chen_icc08, chen_globecom08, chen_eurasip09}, we investigate the impact of frequency diversity on EED in wideband MIMO systems. We will see that frequency diversity benefits wideband systems on EED, though it has no effect on ergodic capacity. We will give an elaborative analysis on the asymptotic optimum expected EED for high SNR. Its lower bound with infinite frequency diversity will be provided. We will see that the tendency of the asymptotic optimum expected EED reflects well the behavior of the optimum expected EED.

Our results can be easily extended to the case of time diversity, a counterpart to frequency diversity. So, it is not surprising to see that our result of the optimum distortion exponent with respect to frequency diversity order is identical to Gunduz and Erkip's result in \cite{gunduz_it08} with respect to time diversity order. However, via introducing the multiplicative optimum distortion factor, we obtain more information on the impact of the diversity order and thus give a more clear guidance on wideband-system design.

Throughout the paper, vectors and matrices are indicated by bold, $\vert \mathbf{A} \vert$ denotes the determinant of matrix $\mathbf{A}$, $\mathbb{E}_x\{\cdot\}$ denotes expectation over the random variable $x$, the superscript $^\dag$ denotes conjugate transpose, and $(a)_n$ denotes $\Gamma(a+n)/\Gamma(a)$.

\section{System Model}
Assume that a continuous-time white Gaussian source $s(t)$ of bandwidth $W_s$ Hz and source power $P_s$ Watts per second is to be transmitted over a frequency-selective block-fading $N_t$-input $N_r$-output channel of bandwidth $W_c$ Hz which can be divided into $L$ independent subchannels of coherence bandwidth $W_b$ Hz, \emph{i.e.}, $W_c = LW_b$ \cite{goldsmith}. Let $\hat{s}(t)$ denote the recovered source at the receiver.

As stated in \cite[pp.\,248-250]{cover}, each subchannel can be represented by the samples taken $1/2W_b$ seconds apart, \emph{i.e.}, each subchannel is used at $2W_b$ channel uses per second as a time-discrete channel. The output of the $l^{\mathrm{th}}$ subchannel for the $t^{\mathrm{th}}$ channel use is
\begin{equation}
\mathbf{y}_{t,l} = \mathbf{H}_{l}\mathbf{x}_{t,l} + \mathbf{n}_{t,l}
\end{equation}
where $\mathbf{x}_{t,l}\in \mathbb{C}^{N_t}$ is the transmitted subband signal satisfying the long-term power constraint $\mathbb{E}[\mathbf{x}_{t,l}^\dag\mathbf{x}_{t,l}] = P_l/2W_b$, $\mathbf{H}_l\in\mathbb{C}^{N_r\times N_t}$ is the subchannel matrix whose elements are distributed as $\mathcal{CN}(0,1)$, $\mathbf{n}_{t,l}\in \mathcal{C}^{N_t}$ is the additive white noise vector whose elements are distributed as $\mathcal{CN}(0,N_0)$, \emph{i.e.}, the noise spectral density is $N_0/2$ Watts per Hz in each dimension of the complex subchannel. The total transmit power is supposed to be $P$ Watts per second, \emph{i.e.}, $\sum_{l=1}^L P_l = P$.

In the case of uncorrelated channel, the elements in $\mathbf{H}_l$ are independent to each other. In the case of spatially correlated channel, we assume the antennas are correlated at the transmitter but not the receiver. The correlation matrix $\bm{\Sigma} = \mathbb{E}(\mathbf{H}_l\mathbf{H}_l^{\dag})$ is supposed to be the same for all subbands and be a full rank matrix with its diagonal all 1's and distinct eigenvalues $\mathbf{\sigma} = [\sigma_1, \sigma_2, \ldots, \sigma_{\mathrm{N}_{\min}}]$, $0 < \sigma_1 < \sigma_2 < \ldots < \sigma_{N_{\min}}$. It can be seen that in the case of uncorrelated channel, $\bm{\Sigma}$ is an identity matrix with $\sigma_1 = \sigma_2 = \ldots = \sigma_{N_{\min}} = 1$.

Under the assumption that the transmitter does not know the subchannel gains, the transmit power is equally allocated to the transmission over each subchannel, \emph{i.e.}, $P_l = P/L$, $l = 1, \ldots, L$. With the noise power $2N_0W_b$ watts per second, the SNR for each subchannel is
\begin{equation}
\rho_l = \frac{\frac{P}{L}}{2N_0W_b} = \frac{P}{N_0W_c}.
\end{equation}
It can be seen that the SNR for each subchannel is equal to the SNR for the whole channel, $\rho = P/N_0W_c$.

The channel is supposed to be perfectly known at the receiver and the transmission system is assumed to be free of outage, \emph{e.g.}, the transmitter knows the instantaneous channel capacity via scalar feedback and does joint source-channel coding to avoid outage accidents. An alternative example of an outage-free system would be analog (continuous-parameter) or approximate-analog (such as hybrid digital-analog \emph{e.g.} in \cite{phamdo2002} and infinite-layer multiplexing \emph{e.g.} in \cite{tian_it2008}) transmission where no outage happens.

\section{Main Results}

\subsection{Optimum EED for any SNR}
The instantaneous channel capacity is the sum of the subchannel capacities
\begin{equation}
R_c = \sum_{l=1}^L R_{b,l} \quad \text{bits per second}. \label{eq:Rc}
\end{equation}
where the the capacity of the $l^{\mathrm{th}}$ subchannel is
\begin{equation}
R_{b,l} = 2W_b\log_2\left\vert \mathbf{I}_{N_r} + \frac{\rho}{N_t} \mathbf{H}_l\mathbf{H}_l^\dag\right\vert \quad \text{bits per second}.
\end{equation}

The source rate of the white Gaussian source $s(t)$ is \cite{shannon48}
\begin{equation}
R_s = W_s\log_2 \frac{P_s}{D} \quad \text{bits per second}
\end{equation}
where $D$ is the mean squared error, \emph{i.e.}, end-to-end distortion (EED)
\begin{equation}
D = \lim_{T\rightarrow \infty}\frac{1}{T}\int_0^T \vert s(t)-\hat{s}(t)\vert^2\mathrm{d}t.
\end{equation}

In terms of Shannon's inequality \cite{shannon} stretched to the block-fading case,
\begin{equation}
R_s \leq R_c,
\end{equation}
the optimum EED is
\begin{equation}
D^*_L(\eta) = P_s\prod_{l=1}^L\left\vert \mathbf{I}_{N_r}+ \frac{\rho}{N_t}\mathbf{H}_l\mathbf{H}_l^\dag\right\vert^{-\frac{2}{L\eta}} \quad \text{watts per second}
\end{equation}
where $\eta$ is the source-to-channel bandwidth ratio (SCBR), $W_s/W_c$.

Thereby, the optimum expected end-to-end distortion is
\begin{equation}
\begin{split}
ED^*_L(\eta) &= \mathbb{E}_{\mathbf{H}_1, \cdots, \mathbf{H}_L}\left(D^*\right) \\
&= P_s \left[\mathbb{E}_{\mathbf{H}}\left(\left\vert \mathbf{I}_{N_r}+ \frac{\rho}{N_t}\mathbf{H}\mathbf{H}^\dag\right\vert^{-\frac{2}{L\eta}}\right)\right]^L
\label{eq:edl}
\end{split}
\end{equation}
where $\mathbf{H}$ denotes a flat Rayleigh fading MIMO channel.

In terms of Jensen's inequality, we have that for $L > 1$,
\begin{equation}
ED^*_L(\eta) < ED^*_1(\eta), \label{ineq:edl}
\end{equation}
which indicates that the frequency diversity improves the optimum expected EED. That is, for transmitting an analog source, a system over a channel with frequency diversity to be exploited can achieve better EED than a system over a channel of the same bandwidth but without frequency diversity. As an interesting counterpart, it is known that the ergodic capacity cannot be improved by increasing the frequency diversity.

The expression (\ref{eq:edl}) can be rewritten as
\begin{equation}
ED^*_L(\eta) =  P_s^{1-L}\left[ED^*_1(L\eta)\right]^L.
\label{eq:edled1}
\end{equation}
When $L=1$, the analytical expression of $ED^*_1(\eta)$ has been given in \cite{chen_eurasip09}. Hence, the analytical expression of $ED^*_L(\eta)$ is straightforward.

Fig.\ref{fig:edsiml} shows the impact of frequency diversity on the optimum expected EED by evaluating (\ref{eq:edl}). The channel is assumed to be uncorrelated, $N_t = 4$, $N_r = 2$, $\eta = 0.2$, $P_s$ = 1. It can be seen that the optimum expected EED $ED^*_L$ is decreasing with the  frequency diversity order $L$ and the effect is obvious in log-log scale when SNR is relatively high.

\subsection{Optimum asymptotic EED for high SNR}
The asymptotic expression of $ED^*_L(\eta)$ in the high SNR regime can be written as
\begin{equation}
ED^*_{L,\mathrm{asy}}(\eta) = \mu^*_{L}(\eta)\rho^{-\Delta^*_L(\eta)}
\label{eq:edlasy}
\end{equation}
with
\begin{align}
& \Delta^*_L(\eta) = \lim_{\rho\rightarrow\infty}\frac{\log ED^*_L(\eta)}{\log\rho}, \\
& \lim_{\rho\rightarrow\infty}\frac{\log\mu^*_L(\eta)}{\log\rho}  = 0
\end{align}

Related to (\ref{eq:edled1}), we obtain that
\begin{align}
\Delta_L^*(\eta) &= L\Delta_1^*(L\eta), \label{eq:expl}\\
\mu^*_L(\eta) &= P_s^{1-L}\mu_1^*(L\eta)^L. \label{eq:facl}
\end{align}

In \cite{chen_eurasip09}, the closed-form expressions of $\Delta_1^*(\eta)$ and $\mu_1^*(\eta)$ for both cases of spatially uncorrelated and correlated channels have been given. For reading convenience, we review the preceding results as follows:

\begin{itemize}
\item (Theorem 2 and Theorem 5 in \cite{chen_eurasip09})

The optimum distortion exponent is
\begin{equation}
\Delta_{1}^*(\eta) = \sum_{k=1}^{N_{\min}}\min\left\{\frac{2}{\eta}, 2k-1+\vert N_t-N_r\vert\right\}
\end{equation}
with $N_{\min} = \min\{N_t, N_r\}$, irrelevant to the spatial correlation.

\item (Theorem 3 and Theorem 6 in \cite{chen_eurasip09})

\begin{figure*}
\begin{align}
\small
\kappa_l(\beta, t, m, n) &=
\begin{cases}
\Gamma(n-m+1)\frac{\Gamma(\beta-n+m-1)}{\Gamma(\beta)}\prod_{k=2}^t \Gamma(k)\Gamma(n-m+k) \frac{\Gamma(\beta-n+m-2k+2)\Gamma(\beta-n+m-2k+1)} {\Gamma(\beta-k+1)\Gamma(\beta-n+m-k+1)}, \quad & t > 1;\\ \Gamma(n-m+1)\frac{\Gamma(\beta-n+m-1)}{\Gamma(\beta)}, \quad & t = 1; \\
1, \quad & t=0.\label{kl}
\end{cases}
\end{align}
\end{figure*}

\normalsize
Define two four-tuple functions, $\kappa_l(\beta, t, m, n)$ as (\ref{kl}) on the top of next page and
\begin{equation}
\kappa_h(\beta, t, m, n) =
\begin{cases}
\prod_{k=1}^t\Gamma(k)\Gamma(n-m-\beta+k), \qquad & t > 0;\\
1, \quad & t = 0, \label{kh}
\end{cases}
\end{equation}
for $\beta \in \mathbb{R}^+$ and $t \in \{0, \mathbb{Z}^+\}$. The optimum distortion factor $\mu^*_{1}(\eta)$ is given as follows:

\begin{itemize}
\item For $2/\eta \in (0, \vert N_t-N_r\vert+1)$, referred to as the \emph{high SCBR regime}, the optimum distortion factor is
\begin{equation}
\begin{split}
\mu^*_1(\eta) &= P_s{N_t}^{\Delta_1^*(\eta)}\kappa_h\left(\frac{2}{\eta}, N_{\min}, N_{\min},N_{\max}\right)\\
&\quad \times \prod_{k=1}^{N_{\min}}\frac{\sigma_k^{-\frac{2}{\eta}}}{\Gamma(N_{\max}-k+1)\Gamma(N_{\min}-k+1)}
\end{split}
\label{eq:fac1}
\end{equation}
with $N_{\max} = \max\{N_t, N_r\}$.
\item For $2/\eta \in (N_t+N_r-1, +\infty)$, referred to as the \emph{low SCBR regime}, the optimum distortion factor is
\begin{equation}
\begin{split}
\mu^*_1(\eta) &= P_s{N_t}^{\Delta^*_1(\eta)}\kappa_l\left(\frac{2}{\eta}, N_{\min}, N_{\min}, N_{\max}\right)\\
&\quad \times \prod_{k=1}^{N_{\min}}\frac{\sigma_k^{-N_{\max}}}{\Gamma(N_{\max}-k+1)\Gamma(N_{\min}-k+1)}.
\end{split}
\end{equation}

\item For $2/\eta \in [\vert N_t-N_r\vert+1, N_t+N_r-1]$, referred to as \emph{moderate SCBR regime}, the optimum distortion factor in the case of uncorrelated channel is given  by (\ref{eq:fac3}); the one in the case of correlated channel is given  by (\ref{eq:fac3b}). The expressions (\ref{eq:fac3}) and (\ref{eq:fac3b}) are on the top of next page, with
\begin{equation}
s = \left\lfloor \frac{\frac{2}{\eta}+1-\vert N_t-N_r\vert}{2}\right\rfloor \label{eq:s}
\end{equation}
and each element of $\mathbf{V}_3(\bm{\sigma})$,
\begin{equation}
v_{3,ij}= \sigma_i^{-\min\{j-1, \frac{2}{\eta}-\vert N_t-N_r\vert-j\}}.
\end{equation}
Note that we have proved in \cite{chen_eurasip09}
\begin{equation}
\lim_{\bm{\Sigma}\rightarrow \mathbf{I}}\mu^*_{1,\mathrm{cor}} = \mu^*_{1,\mathrm{unc}}.
\end{equation}

\begin{figure*}
\begin{equation}
\mu^*_{1, \mathrm{unc}}(\eta) = \begin{cases} P_s{N_t}^{\Delta^*_1(\eta)}\;\frac{\kappa_l(\frac{2}{\eta}, s, N_{\min}, N_{\max})\kappa_h(\frac{2}{\eta}-2s, N_{\min}-s, N_{\min}, N_{\max})}{\prod_{k=1}^{N_{\min}}\Gamma(N_{\max}-k+1)\Gamma(N_{\min}-k+1)}, \quad &\mod\{\frac{2}{\eta}+1-\vert N_t-N_r\vert,2\} \neq 0; \\
P_s{N_t}^{\Delta^*_1(\eta)}\log\rho\;\frac{\kappa_l(\frac{2}{\eta}, s-1, N_{\min}, N_{\max})\kappa_h(\frac{2}{\eta}-2s, N_{\min}-s, N_{\min}, N_{\max})} {\prod_{k=1}^{N_{\min}}\Gamma(N_{\max}-k+1)\Gamma(N_{\min}-k+1)},\quad &\mod\{\frac{2}{\eta}+1-\vert N_t-N_r\vert,2\} = 0
\end{cases}\label{eq:fac3}
\end{equation}
\end{figure*}
\begin{figure*}
\begin{equation}
\mu_{1,\mathrm{cor}}^*(\eta) =
\frac{(-1)^{\frac{s(s-1)}{2}}\vert\mathbf{V}_3(\bm{\sigma})\vert} {\prod_{k=1}^{N_{\min}}\sigma_k^{\vert N_t-N_r\vert+1}\prod_{1\leq m < n \leq N_{\min}}(\sigma_n-\sigma_m)} \prod_{k=1}^{N_{\min}-s}\frac{(k)_s}{(\vert N_t-N_r\vert-\frac{2}{\eta}+s+k)_s}\,\mu^*_{1,\mathrm{unc}}(\eta)\label{eq:fac3b}
\end{equation}
\end{figure*}
\end{itemize}
\end{itemize}

From the above review and the expressions (\ref{eq:expl}) and (\ref{eq:facl}), it can be seen that when the channel is  frequency selective, the ranges of SCBR regime are in terms of $2/L\eta$, instead of $2/\eta$ in the case of flat channel.

Consider the optimum distortion exponent $\Delta_L^*$. When a system is in the low SCBR regime, i.e., $2/L\eta \in (N_t+N_r-1, +\infty)$,
\begin{equation}
\Delta_L^* = LN_tN_r; \label{eq:explscbr}
\end{equation}
when a system is in the high SCBR regime, i.e, $2/L\eta \in (0, \vert N_t-N_r\vert +1)$,
\begin{equation}
\Delta_L^* = 2N_{\min}/\eta; \label{eq:exphscbr}
\end{equation} when a system is in the moderate SCBR regime, i.e., $2/L\eta \in [\vert N_t - N_r\vert +1, N_t+N_r-1]$,
\begin{equation}
\Delta^*_{L} = Ls(s+\vert N_t-N_r\vert)+\frac{2(N_{\min}-s)}{\eta} \label{eq:expmscbr}
\end{equation}
with $s$ given in (\ref{eq:s}).

Related to (\ref{eq:explscbr}), (\ref{eq:exphscbr}) and (\ref{eq:expmscbr}), when a system is in the low or moderate SCBR regime, the optimum distortion exponent $\Delta_L^*$ is monotonically increasing with the frequency diversity order $L$; whereas, when a system is in the high SCBR regime, it has nothing to do with $L$. If a system is in the low SCBR regime when $L=1$, related to the definitions of the SCBR regimes, increasing $L$ continuously will make the system migrate into the moderate SCBR regime and finally into the high SCBR regime. The transit point from the moderate SCBR regime to the high SCBR regime is
\begin{equation}
L^* = \left\lceil \frac{2}{\eta(\vert N_t-N_r\vert+1)}\right\rceil,
\end{equation}
beyond which the increase of frequency diversity has no effect on the optimum distortion exponent, \emph{i.e.}, it does not affect the slope of $ED^*_{\mathrm{L,asy}}$.

When a system is in the high SCBR regime, in terms of (\ref{eq:facl}) and (\ref{eq:fac1}), the optimum distortion factor is
\begin{equation}
\mu^*_L = P_s N_t^{\frac{2N_{\min}}{L\eta}}\left[\prod_{k=1}^{N_{\min}} \frac{\sigma_k^{-\frac{2}{\eta}}\Gamma(\vert N_t-N_r\vert-\frac{2}{L\eta}+k)}{\Gamma(\vert N_t-N_r\vert+k)}\right]^L.
\end{equation}
Let
\begin{equation}
\varphi(L) = \prod_{k=1}^{N_{\min}} \frac{\sigma_k^{-\frac{2}{\eta}}\Gamma(\vert N_t-N_r\vert-\frac{2}{L\eta}+k)}{\Gamma(\vert N_t-N_r\vert+k)}.
\end{equation}
Since $\varphi(L)<1$  and $\frac{\mathrm{d} }{\mathrm{d} L}\varphi(L) > 0$, the derivative of $\mu^*_L$ with respect to $L$
\begin{equation}
\frac{\mathrm{d} }{\mathrm{d} L}\mu_L^* = P_sN_t^{\frac{2N_{\min}}{\eta}}\varphi(L)^L\ln \varphi(L)\cdot\frac{\mathrm{d} }{\mathrm{d} L}\varphi(L) < 0.
\end{equation}
Namely, when a system is in the high SCBR regime, the optimum distortion factor $\mu^*_L$ is monotonically decreasing with $L$.

Synthetically, it is proved that the optimum asymptotic expected EED $ED^*_{L,\mathrm{asy}}$ is monotonically decreasing with the frequency diversity order $L$ and when $L>L^{*}$, only the offset of $ED^*_{L,\mathrm{asy}}$ is impacted by $L$ relative to the case when $L = L^{*}$.

Fig.\ref{fig:edasyl} shows the impact of frequency diversity on optimum asymptotic expected EED $ED^*_{L,\mathrm{asy}}$ by evaluating (\ref{eq:edlasy}) whose closed-form expression is provided. The setting is the same as for Fig.\ref{fig:edsiml}. Relating to Fig.\ref{fig:edsiml}, we can see that the tendency of asymptotic optimum expected EED with respect to frequency diversity reflects the behavior of optimum expected EED. It can be seen that when $L > L^*$, the benefit from increasing frequency diversity is much less than when $L \leq L^*$.

From Fig.\ref{fig:edsiml} and Fig.\ref{fig:edasyl}, we can see that the asymptotic lines with $L = 3$ and $L = 4$ are very close to the curves of $ED_L^*$ when SNR is greater than 20 dB and the asymptotic lines with $L > 4$ are very close to the curves of $ED_L^*$ when the SNR is greater than 15 dB. It illustrates that for practical high SNR, we can use the analysis on the asymptotic EED instead on the EED because
\begin{equation}
ED^*_L \approx ED^*_{L, \mathrm{asy}}.
\end{equation}
Thereby, due to the closed form expression of asymptotic EED, our analysis can be dramatically simplified.

\subsection{EED cannot be vanished by infinite frequency diversity}
It is straightforward that the ergodic capacity is
\begin{equation}
C = 2W_c\mathbb{E}_{\mathbf{H}}\left(\log_2\vert\mathbf{I}+\frac{\rho}{N_t}\mathbf{H}\mathbf{H}^{\dag}\vert\right)
\quad \text{bits per second}
\end{equation}
where $\mathbf{H}$ denotes a flat Rayleigh fading MIMO channel.

A channel with infinite frequency diversity order can be regarded as a fast-fading channel. Thereby, the lower bound on the optimum expected EED $ED^*_L$ related to infinite frequency order is
\begin{equation}
\lim_{L\rightarrow\infty}ED^*_{L} = P_s\,2^{-\frac{2}{\eta}\mathbb{E}_{\mathbf{H}}\left(\log_2\vert\mathbf{I}+\frac{\rho}{N_t}\mathbf{H}\mathbf{H}^{\dag}\vert\right)}.
\end{equation}
In Fig.\ref{fig:edsiml}, the lower bound on $ED^*_L$ is marked by the lowest dash line with circles.

In the following, we will focus on deriving a lower bound on the optimum asymptotic expected EED $ED^*_{L, \mathrm{asy}}$ in closed form.

Since the system is in the high SCBR regime when $L$ goes to infinity, By Lemma \ref{lemma:1} (see Appendix), the optimum distortion factor
\begin{equation}
\lim_{L\rightarrow \infty}\mu^*_L = P_sN_t^{\frac{2N_{\min}}{\eta}}e^{\frac{2\gamma N_{\min}}{\eta} -\frac{2}{\eta}\sum_{k=1}^{N_{\min}}H_{\vert N_t-N_r\vert+k-1}}\prod_{k=1}^{N_{\min}}\sigma_k^{-\frac{2}{\eta}}
\end{equation}
where $\gamma$ is the Euler-Mascheroni constant and $H_n$ is the harmonic number with the order $n$,$H_n = \sum_{k=1}^n \frac{1}{k}$.

Therefore, the lower bound on the optimum asymptotic expected EED $ED^*_{L, \mathrm{asy}}$ is
\begin{equation}
\begin{split}
& \lim_{L\rightarrow\infty}ED^*_{L, \mathrm{asy}} \\
&=  P_sN_t^{\frac{2N_{\min}}{\eta}}e^{\frac{2\gamma N_{\min}}{\eta} -\frac{2}{\eta}\sum_{k=1}^{N_{\min}}H_{\vert N_t-N_r\vert+k-1}}\prod_{k=1}^{N_{\min}}\sigma_k^{-\frac{2}{\eta}}\rho^{-\frac{2N_{\min}}{\eta}}.
\end{split}\label{eq:edlb}
\end{equation}
In Fig.\ref{fig:edasyl}, the lower bound on $ED^*_{L, \mathrm{asy}}$ is marked by dash line.

From Fig.\ref{fig:edsiml} and Fig.\ref{fig:edasyl}, we can see that when $L$ approaches infinite, for SNR $> 10$ dB, the lower bound on the optimum expected EED $ED^*_L$ is almost overlapped by the lower bound on the optimum asymptotic expected EED $ED^*_{L, \mathrm{asy}}$
\begin{equation}
\lim_{L\rightarrow \infty}ED^*_L \approx \lim_{L\rightarrow \infty}ED^*_{L, \mathrm{asy}}.
\end{equation}
That is, for a wideband MIMO system with high frequency diversity, for a rather large range of SNR, the results of an analysis on the asymptotic EED could be used as the results on the EED.

\subsection{Impact of spatial correlation}
In \cite{chen_eurasip09}, we have stated that the effect of spatial correlation on the optimum asymptotic expected EED is only on the optimum distortion factor but not the optimum distortion exponent. The spatial correlation decreases the optimum distortion factor and thus worsen EED.

Since
\begin{equation}
\sum_{k=1}^{N_{\min}}\sigma_k^{-\frac{2}{\eta}} = N_{\min},
\end{equation}
in terms of the inequality between the arithmetic mean and the geometric mean, we have
\begin{equation}
\prod_{k=1}^{N_{\min}}\sigma_k < 1.
\end{equation}
Hence, related to (\ref{eq:edlb}), we have
\begin{equation}
\lim_{L\rightarrow\infty}ED^*_{L, \mathrm{asy, unc}} <  \lim_{L\rightarrow\infty}ED^*_{L, \mathrm{asy, cor}}.
\end{equation}

Fig.\ref{fig:edcorr} shows the impact of spatial correlation on optimum asymptotic expected EED. In this example, we consider a well-known correlation model as in \cite{chiani}: the exponential correlation with $\bm{\Sigma}= \{r^{\vert i-j\vert}\}_{i,j = 1,\cdots, N_{\min}}$ and $r \in (0,1)$ \cite{aalo}.

\section{Conclusion}
We investigated the impact of frequency diversity on the optimum expected end-to-end distortion (EED). It was shown that exploiting frequency diversity can improve EED. Via the analysis  on the optimum asymptotic expected EED, we showed that there is a transit point beyond which the effect of increasing the frequency diversity becomes relatively slight and the EED cannot be vanished by increasing the frequency diversity. Related to infinite frequency diversity, the lower bounds on optimum expected EED and optimum asymptotic expected EED were provided. The impact of spatial correlation was shown as well. Though the results in this paper are derived under assumption of outage-free systems, they can be bounds for outage-suffering systems.

\appendix[]
\newtheorem{lemma}{Lemma}
\begin{lemma}\label{lemma:1}
\begin{equation}
\lim_{L\rightarrow \infty} \left[\frac{\Gamma\left(n-\frac{a}{L}\right)}{\Gamma(n)}\right]^L = e^{a\gamma+\frac{a}{n}-\sum_{k=1}^n\frac{a}{k}} ,\quad n\in \mathbb{N}, a \neq 0
\end{equation}
where $\gamma$ is the Euler-Mascheroni constant.
\end{lemma}
\begin{IEEEproof}
Omitted due to the space limit.
\end{IEEEproof}

\begin{figure}
\centering
\includegraphics[width = 7 cm , height = 6.5 cm]{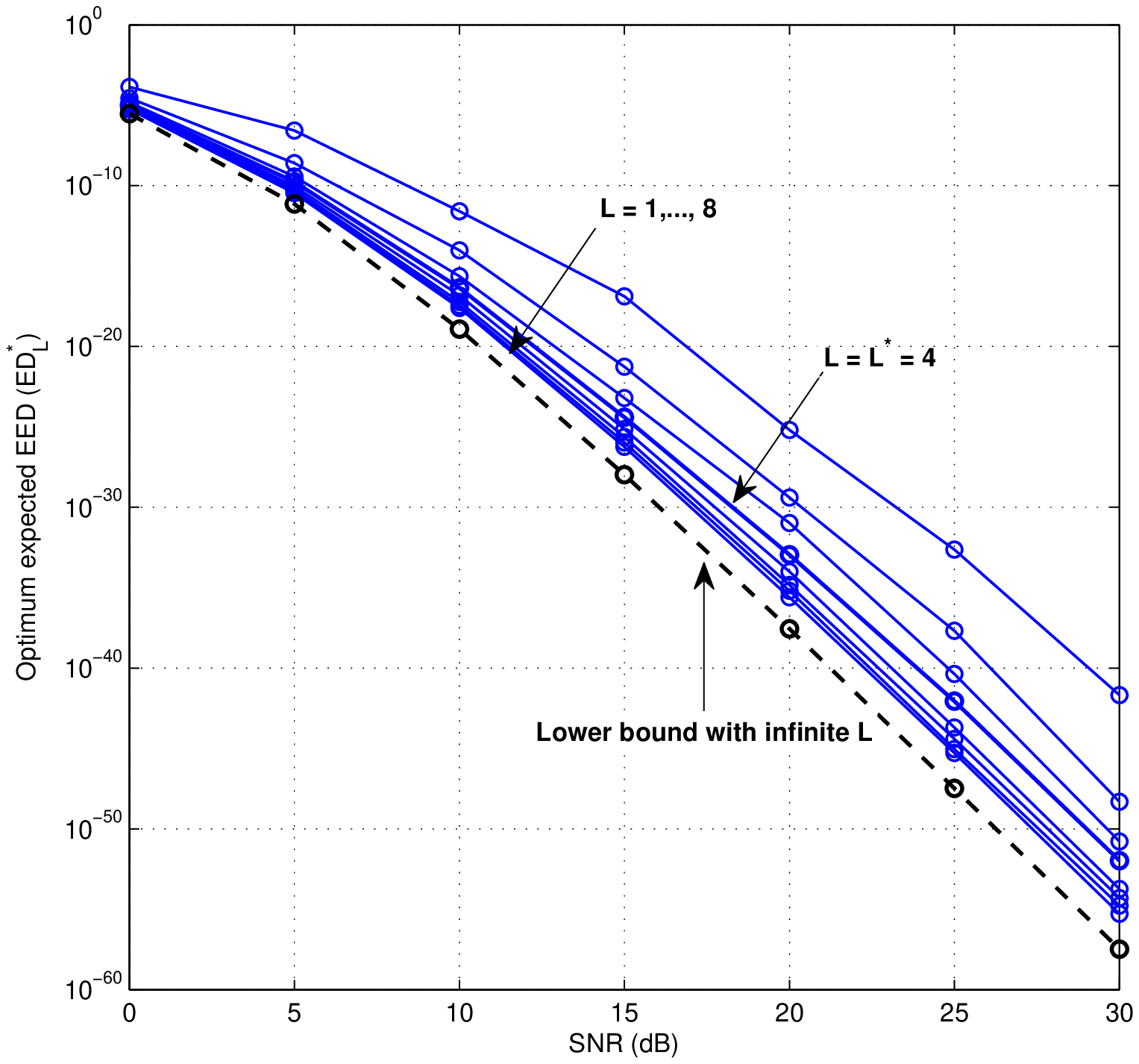}
\caption{Impact of frequency diversity on optimum expected EED.}
\label{fig:edsiml}
\end{figure}

\begin{figure}
\centering
\includegraphics[width = 7cm , height = 6.5cm]{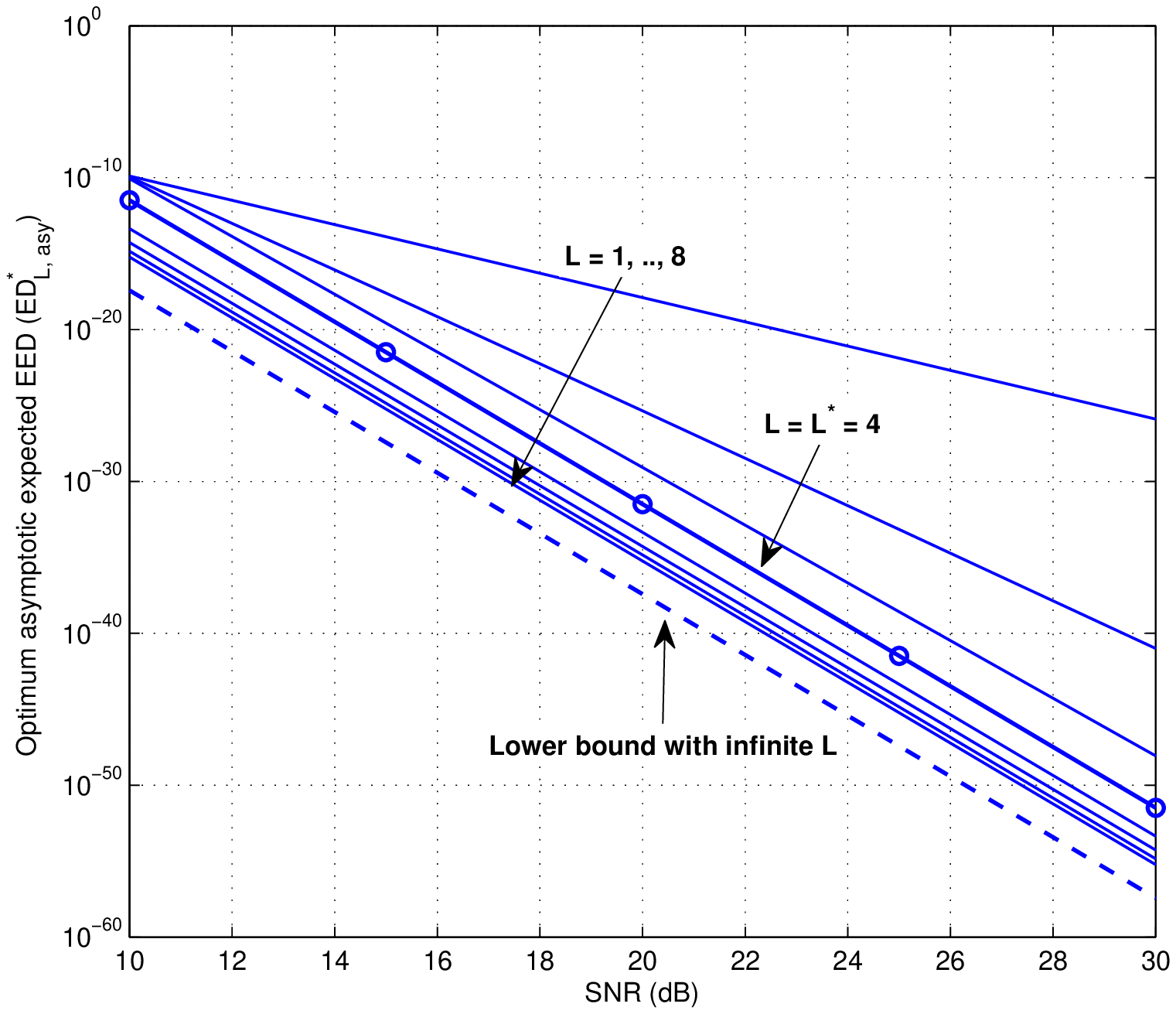}
\caption{Impact of frequency diversity on optimum asymptotic expected EED.}
\label{fig:edasyl}
\end{figure}

\begin{figure}
\centering
\includegraphics[width = 7cm , height = 6.5cm]{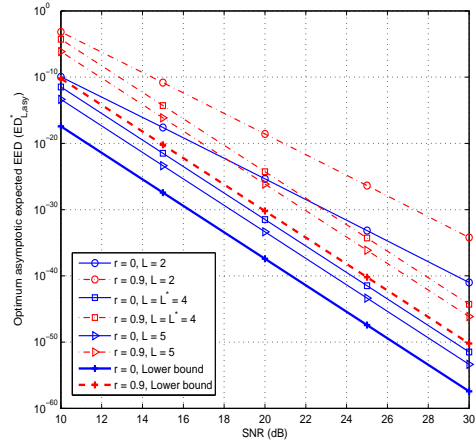}
\caption{Impact of spatial correlation on optimum asymptotic expected EED.}
\label{fig:edcorr}
\end{figure}
\end{document}